\documentclass[conference]{IEEEtran}
\IEEEoverridecommandlockouts
\renewcommand{\thetable}{\arabic{table}}
\usepackage[utf8]{inputenc}
\usepackage{booktabs}
\usepackage{caption}
\usepackage{url}
\usepackage{graphicx}
\usepackage{authblk} 

\usepackage{subcaption} 
\usepackage{cite}
\usepackage{amsmath,amssymb,amsfonts}
\usepackage{algorithmic}
\usepackage{graphicx}
\usepackage{textcomp}
\usepackage{xcolor}
\usepackage[linesnumbered, ruled, vlined]{algorithm2e}
\usepackage{wasysym}
\usepackage{multirow}
\usepackage{graphicx}
\usepackage{xcolor}
\usepackage{pdfpages}
\usepackage{colortbl}
\usepackage{ragged2e}
\usepackage{pifont}
\usepackage{booktabs}
\usepackage{amsmath}
\usepackage[T1]{fontenc}   
\usepackage[utf8]{inputenc}
\def\BibTeX{{\rm B\kern-.05em{\sc i\kern-.025em b}\kern-.08em
    T\kern-.1667em\lower.7ex\hbox{E}\kern-.125emX}}
\begin{document}

\newboolean{showcomments}
\setboolean{showcomments}{true}
\ifthenelse{\boolean{showcomments}}
{

\newcommand{\mynote}[2]{
    {\color{#1}Note: #2}
}

\newcommand{\note}[2]{
    {\color{#1}Xinjiao: #2}
}

\newcommand{\Dirk}[2]{
    {\color{#1}\fbox{\begin{minipage}{0.8\linewidth}Dirk: #2\end{minipage}}}
}

\newcommand{\Yisu}[2]{
    {\color{#1}\fbox{\begin{minipage}{0.8\linewidth}Yisu: #2\end{minipage}}}
}

} 
{ 
\newcommand{\mynote}[2]{}
\newcommand{\note}[2]{}
\newcommand{\Dirk}[2]{}
\newcommand{\Yisu}[2]{}
\newcommand{\Xinjiao}[2]{}
} 

\newcommand{\xinjiao}[1]{\note{purple}{#1}}
\newcommand{\dirk}[1]{\Dirk{blue}{#1}}
\newcommand{\yisu}[1]{\Yisu{red}{#1}}

\title{PacTrain: Pruning and Adaptive Sparse Gradient Compression for Efficient Collective Communication in Distributed Deep Learning}


\author{
Yisu Wang, Ruilong Wu, Xinjiao Li, Dirk Kutscher\textsuperscript{*} \\
Hong Kong University of Science and Technology (Guangzhou) \\
\textsuperscript{*} Corresponding author is Dirk Kutscher, dku@hkust-gz.edu.cn
}

\maketitle

\begin{abstract}

Large-scale deep neural networks (DNN) exhibit excellent performance for various tasks. As DNNs and datasets grow, distributed training becomes extremely time-consuming and demands larger clusters. A main bottleneck is the resulting gradient aggregation overhead. While gradient compression and sparse collective communication techniques are commonly employed to alleviate network load, many gradient compression schemes do not achieve acceleration of the training process while also preserving accuracy. This paper introduces PacTrain, a novel framework that accelerates distributed training by combining pruning with sparse gradient compression. Active pruning of the neural network makes the model weights and gradients sparse. By ensuring the global knowledge of the gradient sparsity among all distributed training workers, we can perform lightweight compression communication without harming accuracy. We show that the PacTrain compression scheme achieves a near-optimal compression strategy while remaining compatible with the all-reduce primitive. Experimental evaluations show that PacTrain improves training throughput by 1.25 to 8.72$\times$ compared to state-of-the-art compression-enabled systems for representative vision and language models training tasks under bandwidth-constrained conditions.

\end{abstract}

\section{Introduction}
\label{sec:intro}


Over the past decade, deep learning (DL) has seen significant growth driven by the development of increasingly complex large-scale models and massive datasets \cite{LaMDA, sigcomm22-chen, netllm}. These DL models exhibit excellent performance in various computer vision (CV) and neutral language processing (NLP) tasks \cite{touvron2023llama, gpt3, vision-transformme,cv-task-ex1}. However, their immense scale makes large-scale distributed training (DT) time-consuming and expensive \cite{StellaTrain}.

Large-scale DT 
incurs a high communication overhead, which is a challenge that spans large and small-scale setups. In large clusters with tens of thousands of GPUs provided by cloud providers (\textit{e.g.}, AWS, Azure, Alibaba Cloud) \cite{crux, megascale}, communication overhead increases linearly as the number of workers increases.
Similarly, non-hyperscaler deployments, such as single- or multi-compute clusters in university labs with consumer-grade GPUs, often face communication bottlenecks owing to the constrained and highly variable bandwidths in such systems \cite{StellaTrain}. 

Gradient compression (GC) has emerged as a promising solution for mitigating this communication bottleneck by reducing the volume of the exchanged gradient data \cite{hotnets24-gredient-compression, quant-example1, terngrad, thc, topk, dgc, powersgd, grace}. Most compression schemes (\textit{e.g., }, TopK \cite{topk}, TernGrad \cite{terngrad}, THC \cite{thc}, and DGC \cite{dgc}) are lossy and introduce some compression errors, meaning they often compromise both the convergence rate and attainable accuracy, and thus  fail to accelerate the training process. 

Another potential approach for addressing communication bottlenecks in DDL is sparse collective communication (SCC). In recent years, practitioners in the deep learning community have observed that gradient sparsity often occurs in DNN training and have sought to exploit it \cite{omnireduce, zen, EmbRace, espresso-eurosys23}. For instance, OmniReduce \cite{omnireduce} introduces a streaming aggregation to transfer sparse gradient blocks, and Zen \cite{zen} presents a hashing algorithm that leverages parallel GPU computing to achieve balanced communications. However, these approaches assume that gradients are sparse, which may not always be the case in practice. 

Nevertheless, as pointed out by prior studies \cite{hotnets24-gredient-compression, grace}, a common problem of GC and SCC is that most schemes (\textit{e.g.}, DGC, OmniReduce, and Zen) are not compatible with all-reduce. All-reduce \cite{mccs, swing} is a highly scalable collective communication operation used to aggregate gradients among a set of DT workers. Another widely used gradient collection method is the parameter server (PS) that performs gradient collection on a centralized server \cite{ps-limu}. All-reduce is inherently more parallel and scalable than PS based aggregation \cite{hotnets24-gredient-compression}.

To address these challenges, we present PacTrain, a communication-efficient distributed training framework that combines model pruning and adaptive gradient compression. PacTrain can actively change the training gradient distribution through an unstructured pruning technique and leverages an efficient, non-lossy gradient compression scheme that is compatible with the all-reduce primitive to exploit gradient sparsity.

However, transforming this high-level idea into an existing DT framework, such as PyTorch Distributed Data Parallel (DDP) \cite{pytorch2} is not straightforward because of the abstracted communication interface. DDP is a widely used framework that accelerates training by distributing model replicas across multiple GPUs, synchronizing gradients, and averaging them to ensure consistency. Most DT frameworks, including DDP, offer communication hook functions that enable developers to customize gradient synchronization algorithms. However, these hooks typically expose only a reformatted one-dimensional gradient tensor, where gradient names are removed and gradient orders are rearranged. This abstraction complicates the establishment of a direct correspondence between the reformatted gradient tensor and the original model weights. Re-implementing the distributed training mechanism from scratch risks loses the performance optimizations inherent to the native frameworks. To address this, we propose a mask tracker mechanism that captures sparsity patterns within the reformatted gradient tensor, enabling efficient alignment and customization.

By ensuring that all distributed training workers share global knowledge of gradient sparsity, we can use a pruning mask to reformat sparse gradients into low-dimensional dense gradient tensors. This approach enables lightweight gradient compression without compromising accuracy during the training process. Tab.~\ref{tab:compare} summarizes how the different methods of GC and SCC impact one or both factors that determine the Time-To-Accuracy (TTA): the iteration time and convergence speed compared with PacTrain. Compared to these methods, PacTrain can provide both fast convergence and compatibility with all-reduce, and it can improve TTA performance.


We conducted extensive experiments on two datasets: CIFAR-10, CIFAR-100 \cite{cifar10}, using VGG19 \cite{vgg}, ResNet18, ResNet152 \cite{resnet152}, and ViT-Base-16 \cite{vision-transformme} models. Our evaluations include specific link speeds of 100 Mbps, 500 Mbps, and 1 Gbps, chosen to reflect the WAN bandwidth in small multicluster scenarios. Our evaluation shows that PacTrain effectively minimizes TTA under bandwidth-constrained conditions. Our implementation demonstrates that PacTrain successfully adapts the compression scheme to prune-aware-distributed training scenarios and reduces TTA by up to 8.72$\times$ compared with native PyTorch DDP with variable-constrained network bandwidth.

\begin{table}[t]
\caption{Impact of acceleration methods on training metrics.}
\centering
\renewcommand{\arraystretch}{1.5} 
\begin{tabular}{>{\raggedright\arraybackslash}p{3cm}ccc}
\toprule
\textbf{Method} & \textbf{Conv. Speed} & \textbf{Compatibility} & \textbf{TTA} \\
\midrule
\rowcolor{gray!20} PacTrain & \ding{51} & \ding{51} & \ding{51} \\
THC & \ding{51} & \ding{55} & \ding{51} \\
Terngrad & \ding{55} & \ding{51} & ? \\
DGC & \ding{55} & \ding{51} & ? \\
OmniReduce & \ding{51} & \ding{55} & \ding{51} \\
Zen & \ding{51} & \ding{55} & \ding{51} \\
\bottomrule
\end{tabular}
\vspace{0.5em} 

\justifying 
\footnotesize Each method may positively (marked \ding{51}) or negatively (marked \ding{55}) affect performance. The performance of the methods marked with ? will vary according to the choice of the model architecture and training dataset. Compatibility denotes whether the method is compatible with all-reduce operation.
\label{tab:compare}
\end{table}

This study makes the following contributions.

\textit{Contribution 1.}  We propose that pruning can be used to enhance the gradient compression, thereby accelerating the process. Traditional pruning after training drops a subset of weights to accelerate inference and save memory size, but we show that the gradient sparsity enforcement (GSE) method based on pruning is crucial for realizing efficient distributed deep learning.

\textit{Contribution 2.} We propose 
that all distributed training workers share global knowledge of gradient sparsity and propose a highly efficient, non-lossy compression scheme based on this property that is compatible with all-reduce operations. Our framework can be easily integrated into existing DT frameworks with minimal effort.

\textit{Contribution 3.} Our comprehensive experiments on large-scale DL training tasks demonstrate that PacTrain outperforms traditional GC methods and achieves state-of-the-art TTA performance.

The rest of this paper is structured as follows: Section
\ref{sec:related work} discusses related
work. Section \ref{sec:design}
describes the system design and how to change the gradient distribution via pruning, and Section \ref{sec:eval}
provides a detailed evaluation of different machine learning models
in a real-world network, and Section \ref{sec:conclusion} concludes the paper with a summary of our insights and a perspective on future work.


\section{Related Work}
\label{sec:related work}
\subsection{Gradient Aggregation}

Data parallelism is widely used for training large-scale DL models.  In data-parallel DDL, each GPU worker holds a replica of the DNN model, and the training dataset is partitioned among the workers. Each GPU worker independently performs \textit{forward and backward propagation} on its partition, computing gradients that are aggregated among GPUs.

To mitigate the communication tension with training distributed DDL, many GC and SCC algorithms have been proposed in the community. Techniques such as quantization \cite{thc}, sparsification \cite{dgc}, and sparse collective communication \cite{zen, omnireduce} compress gradients to reduce the amount of data transmitted. However, as previously mentioned, existing GC and SCC methods often lack compatibility with all-reduce collectives. The primary challenge lies in avoiding decompression and recompression at each worker, as this could introduce substantial computational overhead or, in some cases, accumulate compression-induced errors, potentially impacting the training process.

\subsection{Neural Network Pruning}

Neural network pruning is a sparsification technique that reduces the memory cost and computation load by removing less important weights. The importance of model weights can be evaluated using various criteria, such as the absolute value of the weights \cite{han2015pruning} and the L1/L2 norm of the layers \cite{pruning_conv}. Conventional pruning techniques are mainly applied in a train-prune-fine-tune manner \cite{pruning4inference} to improve efficiency at the inference stage.

\textbf{Pruning-Aware-Training.} Conventional exploration of pruning during the training reduced training computation cost at each iteration but suffered noticeable performance degradation. Recently, researchers have proposed several policies to prune as early as possible during training without hurting performance such as the Lottery Ticket Hypothesis (LTH) \cite{2019lottery, chen2021earlybert} and Early Pruning Indicator \cite{when2prune}. 

LTH shows that the existence of sparse sub-networks at initialization can achieve almost the same performance as the dense model when trained alone. By identifying these optimal sparse sub-networks after several training epochs, we can prune most of the model weights and reduce both the computation and communication costs during training. 

The effectiveness of these strategies is also based on the empirical observation that network representations remain similar after the early phase of training \cite{kornblith2019similarity}. Once the model stabilizes, only a small portion of the weights in the network is critical to its performance. Therefore, we can perform targeted training without the need to compute the gradients for all neurons \cite{jia2024modelsparsitysimplifymachine}. 

\textbf{Pruning-Aware-Distributed-Training. } When we know in advance that the model weights are sparse, we can leverage this sparsity to customize our specific compression algorithm based on the sparse distribution of the model, which leads us to optimize the precision of the compression techniques and further reduce the gradient synchronization overhead. 

Recent studies \cite{deep-zero, FedMef, ZeroFL, FedDST, FedTiny} accelerate distributed learning through dynamic sparse training. For example, DeepZero \cite{deep-zero} employs pruning in zeroth-order optimization to induce gradient sparsity and improve training efficiency. FedMef \cite{FedMef}, FedDST \cite{FedDST}, and FedTiny \cite{FedTiny} perform pruning and regrowth on devices to expedite federated learning (FL). However, pruning typically can achieve only 30\% to 80\% sparsity, which is insufficient for OmniReduce’s compression algorithm which requires approximately 1\% sparsity. Using compression schemes like COO (Coordinate List)) also increases heavy computational overhead. Unfortunately, these studies do not detail how to efficiently compress and transmit the sparse gradients.

\begin{figure}
{
\includegraphics[width=\linewidth]{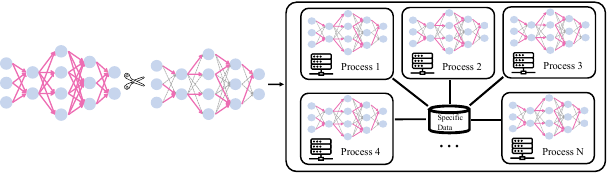}}
\caption{Pruning and Fine-tune on specific data.}
\label{fig:fig1c}
\end{figure}

\section{PacTrain Design}
\label{sec:design}

This section first outlines the problem setup and describes the system design of PacTrain. We then explain how pruning changes the gradient distribution and how the gradient binarization technique preserves the accuracy impacted by pruning.

\subsection{Problem Setup}
Researchers typically start with a model that has been pre-trained on a general-purpose dataset such as ImageNet \cite{imagenet1K}. Then they will re-train this model on a domain-specific dataset to meet the requirements of a specific task. To speed up training, the dataset is distributed across GPUs, with DDP used to enhance training efficiency. However, due to network bottlenecks, gradient aggregation consumes a significant portion of the total training time.

Formally, we consider gradient aggregation phase as:

\begin{equation}
\label{eq:gradient_aggregation}
X^{t+1} = X^t - \eta \cdot \frac{1}{n}\sum_{i=1}^{n} \Delta(X^t, D_i^t)
\end{equation}

where $X^t$ represents the model weights at iteration $t$, $n$ is the number of training workers, $ \eta $ represents the learning rate, $D_i^t$ denotes the dataset partition assigned to the $i$-th GPU, and $\Delta(X^t, D_i^t)$ are the gradients computed by the $i$-th GPU based on its local data. Gradient aggregation ensures all GPU workers have the same updated model weights $X^{t+1}$ for the next iteration.


Our objective is to derive a specialized sparse model with a mask $m$ during the training process, aiming to optimize the TTA in DDP. Pruning a pre-trained model can lead to some accuracy loss, which depends on the pruning amount. The challenge in our optimization is how to enhance the communication speed in the gradient aggregation phase while maintaining the final test accuracy.
\subsection{Design Overview}
The design of PacTrain is motivated by the limitations of re-training and post-finetuning large models within bottleneck network link scenarios. Previous gradient sparsification techniques \cite{topk, dgc} have two main limitations: computational overhead and incompatibility with all-reduce operations, limiting efficient gradient synchronization \cite{hotnets24-gredient-compression}.

\begin{figure}[t]
\centering
\includegraphics[width=0.8\linewidth]{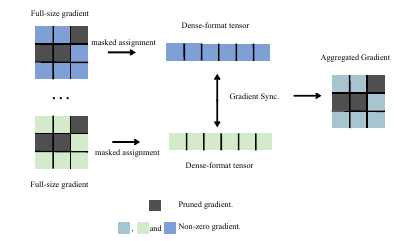}
\caption{Acceleration with masked assignment.}
\label{fig:fig1d}
\end{figure}


PacTrain provides less computational overhead and is compatible with all-reduce to preserve accuracy. To that end, we first train the model on a generic, smaller dataset like CIFAR-10 or TinyImageNet \cite{tinyImgNet}, or use an existing pre-trained model. The pre-trained model contains important weight information for the task, allowing us to perform pruning on the large model, as shown in Fig.~\ref{fig:fig1c}. We then conduct distributed training on our specific dataset to accelerate convergence.


Fig.~\ref{fig:fig1d} shows how PacTrain implements a local sparsification technique. Instead of selecting TopK gradient coordinates, PacTrain selects non-zero gradient elements based on the local weight magnitudes, which remain the same across the distributed workers. The sparse gradients are reformatted into dense tensors to enable all-reduce synchronization, achieving significant compression while maintaining computational efficiency.

\subsection{Adaptive Compression Based on
Pruning-Aware-Distributed-Training}

As illustrated in Fig.\ref{fig:fig1c} and Fig.\ref{fig:fig1d}, the design of PacTrain involves: a pre-trained model is provided to each distributed worker, which then performs task-specific fine-tuning on its subset of data. An unstructured pruning step ensures that the model weights are sparse, optimizing memory and computation efficiency. Subsequently, each worker engages in model update aggregation through the all-reduce operation.

\textbf{Gradient Sparsity Enforcement (GSE).}  To reduce non-zero values in the neural network tensor, a pruning operation is necessary. However, during training, these weights can be updated to non-zero values through non-zero gradients. In other words, the distribution of the gradients is different from the distribution of the model weights. We can explicitly set the corresponding gradients to zero for model weights that are zero during the optimization process. This method ensures gradient sparsity. Specifically, this can be expressed using the following GSE formula:

\begin{equation}
\text{Gradient} = (\text{Weight} \neq 0) \odot \text{Gradient}
\end{equation}

where $\odot$ denotes element-wise multiplication. By repeating this process in every training epoch, the model can progressively adapt to the task while maintaining synchronization in distributed environments.

Our algorithm leverages both sparse gradients and gradient compression during synchronization to significantly reduce communication and computation overhead. Unlike methods that rely solely on gradient magnitudes for compression, this algorithm incorporates the importance of model weights. It effectively eliminates low-magnitude gradients and aligns the gradient update process with both the pruning strategy and intrinsic gradient properties. This design improves data efficiency in constrained networks, accelerates convergence, and enhances the robustness of the model.

The detailed steps of the algorithm are as follows:

\begin{enumerate}
    \item \textbf{Initialization}: Start with a pre-trained model and apply pruning using the specified pruning ratio to create a sparse network.
    \item \textbf{Training}: During each training epoch, apply GSE to ensure that only gradients corresponding to non-zero weights are updated.
    \item \textbf{Sparse Mask Update}: Use a mask tracker to monitor the distribution of sparse weights. When the mask stabilizes, compress the gradient updates to include only non-zero elements; otherwise, perform full synchronization to ensure correctness.
    \item \textbf{Synchronization}: In the distributed environment, communicate only the non-zero elements of the sparse tensor to reduce communication overhead. For unstable masks, use full synchronization to maintain accuracy.
\end{enumerate}


\begin{algorithm}[t]
\caption{Worker Algorithms: Pruning Neural Networks and Send Sparse Tensor during AllReduce Operation}
\label{alg:alg2}
\KwData{
    Pre-trained model: The initial neural network model, \\
    Number of epochs $n$: Total number of training epochs, \\
    $pruning\_ratio$: Ratio used for pruning the model, \\
    $Configuration$: Configuration settings for distributed training, \\
}
\KwResult{Trained sparse model}
\BlankLine
Initialize \textit{Configuration} to find peers and aggregator\;
Prune(model, $pruning\_ratio$) \;
\For{$i = 0 \;  \KwTo \;  n \;  \text{epochs}$}{
            
    Perform training on the model\;
    Apply GSE to the model gradient\;
    \ForEach{$update \in model\_updates$}{
        
        Update sparse mask using Mask Tracker\; 
        \eIf{mask is stable}{
            $compressed\_update \gets compress(update)$\;
            Communicate only non-zero elements in $compressed\_update$;
        }{
             Perform All-reduce on $update$\;
         }
    }
}
\end{algorithm}

\begin{figure*}[ht]
    \centering
    \begin{subfigure}[t]{0.325\linewidth}
        \centering
        \includegraphics[width=\linewidth]{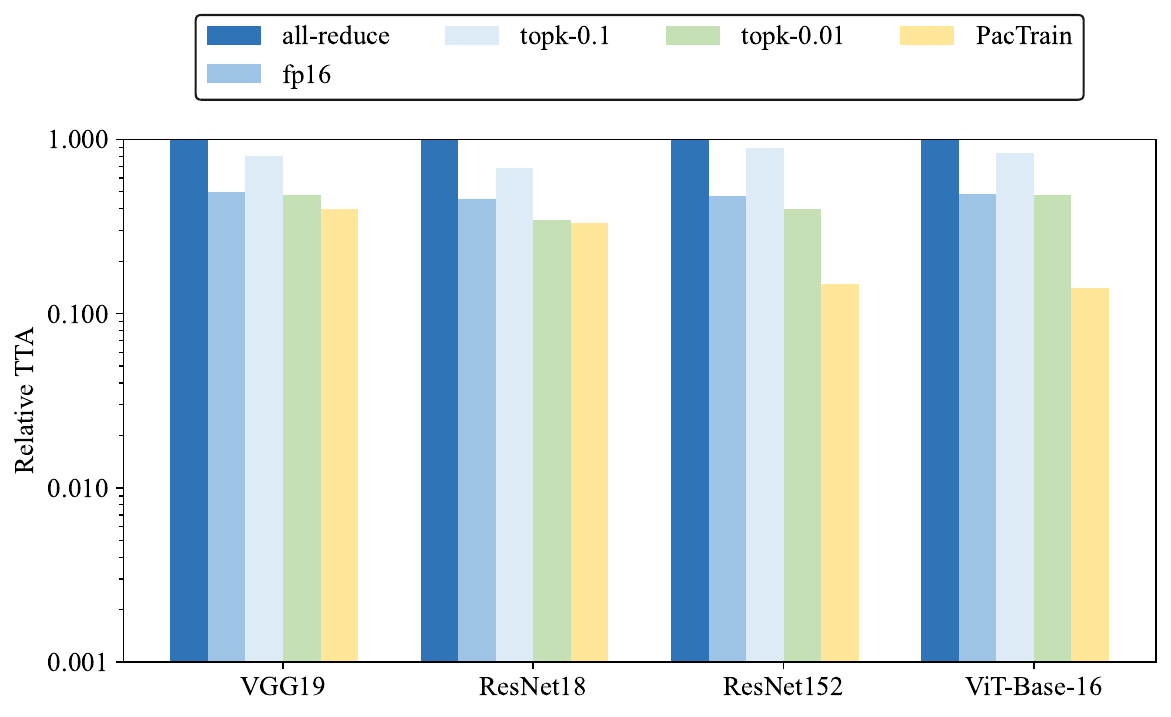}
        \caption{WAN bandwidth = 100 Mbps}
        \label{fig:100M}
    \end{subfigure}
    \hfill
    \begin{subfigure}[t]{0.325\linewidth}
        \centering
        \includegraphics[width=\linewidth]{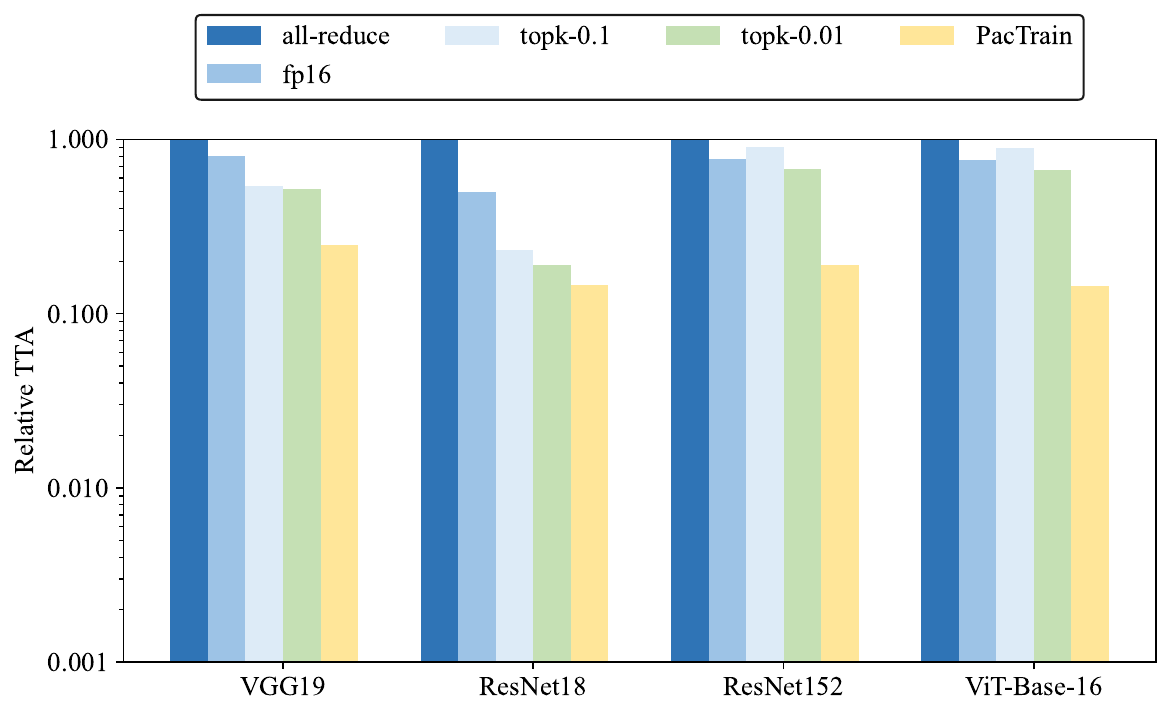}
        \caption{WAN bandwidth = 500 Mbps}
        \label{fig:500M}
    \end{subfigure}
        \begin{subfigure}[t]{0.325\linewidth}
        \centering
        \includegraphics[width=\linewidth]{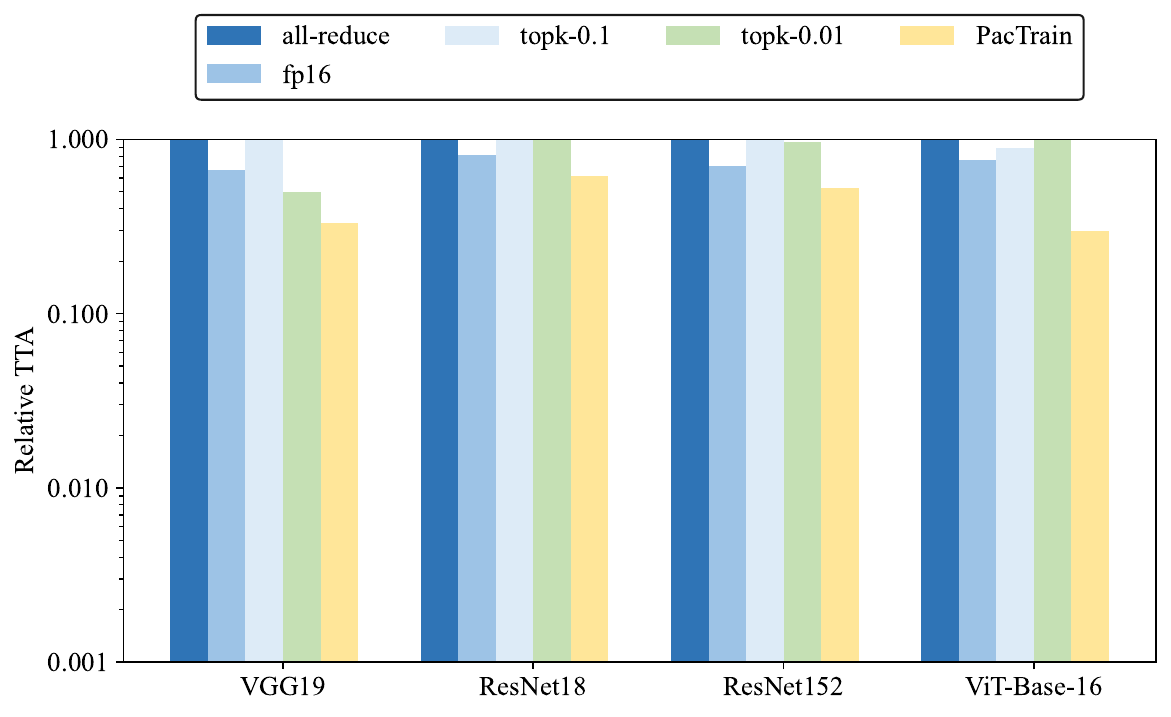}
        \caption{WAN bandwidth = 1 Gbps}
        \label{fig:1G}
    \end{subfigure}
    \caption{End-to-End TTA speedup with different WAN bandwidths (relative to native all-reduce, log scale).}
    \label{fig:TTA comparison}
\end{figure*}

\textbf{Mask Tracker mechanism.} For each training epoch, the worker performs training on the sparse model. We introduce Mask Tracker (MT) to capture the sparsity patterns of the reformatted one-dimensional gradient tensor by maintaining the mapping of the gradient and the model weights. 
Existing DDP frameworks abstract the communication interface to the bucket and gradients are transformed into one-dimensional tensors during synchronization. This transformation removes the original names and order information of the gradients, making it challenging to establish a correspondence between the reformatted gradient tensors and the model weights. 

The Mask Tracker monitors the sparsity patterns of the gradients without the need to redesign the underlying distributed training mechanism. By comparing this mask with the one from the previous iteration, we detect any changes in the sparsity pattern. If the mask remains unchanged over multiple iterations, MT deems the sparsity pattern stable. Once a pattern is considered stable, we communicate only the non-zero (non-masked) portions of the gradient tensor during synchronization. Because the MT operates directly on the flattened gradient tensors without modifying their values, it remains fully compatible with native floating-point AllReduce operations. This compatibility allows us to utilize existing optimizations and communication protocols without alteration.


\subsection{Pruning with Gradient Quantization}

To further reduce communication overhead while preserving final test accuracy, we introduce ternary gradient quantization on top of importance-based pruning during DT. Our insight is applying Terngrad quantization to a pruned model can still direct optimization toward a desirable solution, as the pruning step already retains only the most important weight parameters. Specifically, in PacTrain, we use $ ternarize(\cdot) $ to randomly quantize the gradient $g_t$  into a ternary vector with values $ \in \{-1, 0, +1\} $.

Pruning establishes a fixed mask that permanently zeros out gradients for less important parameters, thus confining the optimization process to a reduced parameter subspace. Although this can potentially slow down convergence or yield suboptimal solutions in non-convex neural networks, ternary quantization introduces controlled randomness. This stochastic element aids the model in escaping local minima.

\textbf{Convergence Analysis.} For each gradient component, the expected value of the quantized gradient is:

\begin{equation}
E(\tilde{g}_t) = E[s \cdot \text{sign}(g_t)] = s \cdot E(\text{sign}(g_t))
\end{equation}

where \( s \) is a scaling factor ensuring that the quantized gradient remains unbiased in expectation. However, quantization increases variance, which may cause fluctuations in the optimization direction and impact convergence speed.

We employ GraSP\cite{grasp} to compute pruning scores that reflect each parameter’s contribution to the gradient flow. Specifically, the pruning score \( S \) is defined as:

\begin{equation}
S = -\theta \odot (H \nabla l(\theta))
\end{equation}

where \( \nabla l(\theta) \) is the gradient of the loss function, and \( H \) is its Hessian matrix. This score allows us to retain parameters critical for maintaining essential gradient directions, thereby enhancing robustness.

In practice, determining the pruning mask based on weight ranking reduces sensitivity to NMSE (Normalized Mean Squared Error, $\text{NMSE}(x, \hat{x}) = \frac{\|x - \hat{x}\|^2}{\|x\|^2}$ ), further improving robustness in distributed training. This ensures that the remaining parameters retain the key gradients necessary for stable optimization.




\section{Evaluation}
\label{sec:eval}

PacTrain is designed to operate in arbitrary network settings, ranging from datacenter networks to the public Internet. We implemented and evaluated PacTrain using a virtual testbed hosted in our laboratory.

\begin{figure}
    \centering
    \includegraphics[width=0.7\linewidth]{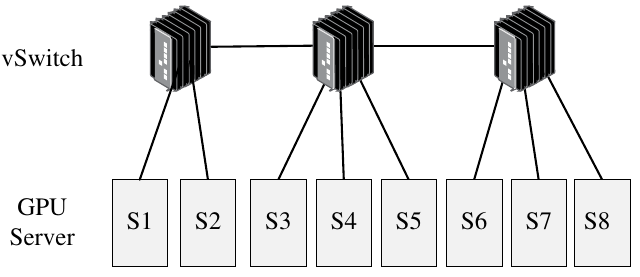}
    \caption{Evaluation topology.}
    \label{fig:topology}
\end{figure}

\subsection{Setup and workloads}
We built our test bed using the ESXI virtualization platform with a server equipped with eight A40 GPUs and an Intel Xeon Platinum 8358P processor with 64 CPU cores. To simulate the network environment shown in Fig.~\ref{fig:topology}, we added three virtual switches, each connected to virtual machines with GPUs.

We developed the PacTrain prototype on top of the PyTorch distributed framework \cite{pytorch2}, implementing distributed machine learning communication, where each GPU performs a learning task and communicates with others using NCCL \cite{nccl} over TCP for gradient aggregation. In our implementation, we utilized PyTorch's
\textit{DistributedDataParallel} communication hook to override the default
all-reduce operation, allowing fine-grained control over how gradients
are communicated across workers.

\textbf{Workloads. } We evaluated our PacTrain prototype by training popular computer vision models on an eight-worker testbed. The models and datasets used in our evaluation included VGG19 \cite{vgg}, ResNet18, ResNet152 \cite{resnet152} and Vit-Base-16 \cite{vision-transformme}, all of which were trained on the CIFAR-100 and CIFAR-10 dataset for 100 epochs. Unless otherwise noted, we set the per-GPU batch size to 32.

\textbf{Metrics. } We evaluate time-to-accuracy (TTA) across all training tasks under varying bandwidth conditions. We define convergence time as the duration required for the model to reach a target accuracy threshold, signifying that the model has fully converged.
    
\subsection{Evaluation Scenarios}
We conducted experiments under three different scenarios using the topology shown in Fig.~\ref{fig:topology}, creating network performance bottlenecks by adjusting the link bandwidths of the two connections to the switch.
%

We configured different bottleneck bandwidths of 100 Mbps, 500 Mbps, and 1 Gbps, representing typical network conditions in cloud environments and small datacenters.

\subsection{End-to-End Performance}

\subsubsection{Time-to-accuracy}
We evaluated PacTrain against FP16 quantization, TopK compression (0.1 rate and 0.01), and AllReduce for VGG19, ResNet18, ResNet152, and Vit-Base-16 models in bandwidth-constrained environments (100 Mbps, 500 Mbps, and 1 Gbps). As most gradient compression algorithms perform similarly to FP16 \cite{hotnets24-gredient-compression}, we selected FP16 and the commonly used TopK algorithms for comparison. Under extremely low bandwidth conditions, PacTrain demonstrates superior convergence speed compared to FP16 and TopK. Unless otherwise specified, we use a compression scheme with a pruning ratio of 0.5.

Fig.~\ref{fig:TTA comparison} shows the end-to-end training time required to reach the same target accuracy in multiple bandwidth-constrained environments, normalized to that of the native all-reduce operation. The experimental results show that, even under constrained network bandwidth, PacTrain can continue training effectively, leading to faster convergence. Compared with methods that do not involve compression, PacTrain achieves a higher TTA reduction by up to 8.72$\times$. Compared to solutions that support gradient compression and quantization (FP16 and TopK), PacTrain achieves TTA reductions of 1.25-7.05$\times$. 

TopK’s gradient exchange relies on all-gather, causing network congestion at a 0.1 compression rate, while a 0.01 rate results in too much information loss, making convergence difficult.

\subsubsection{Iteration and convergence speed}

\begin{figure}
    \centering
    \includegraphics[width=0.8\linewidth]{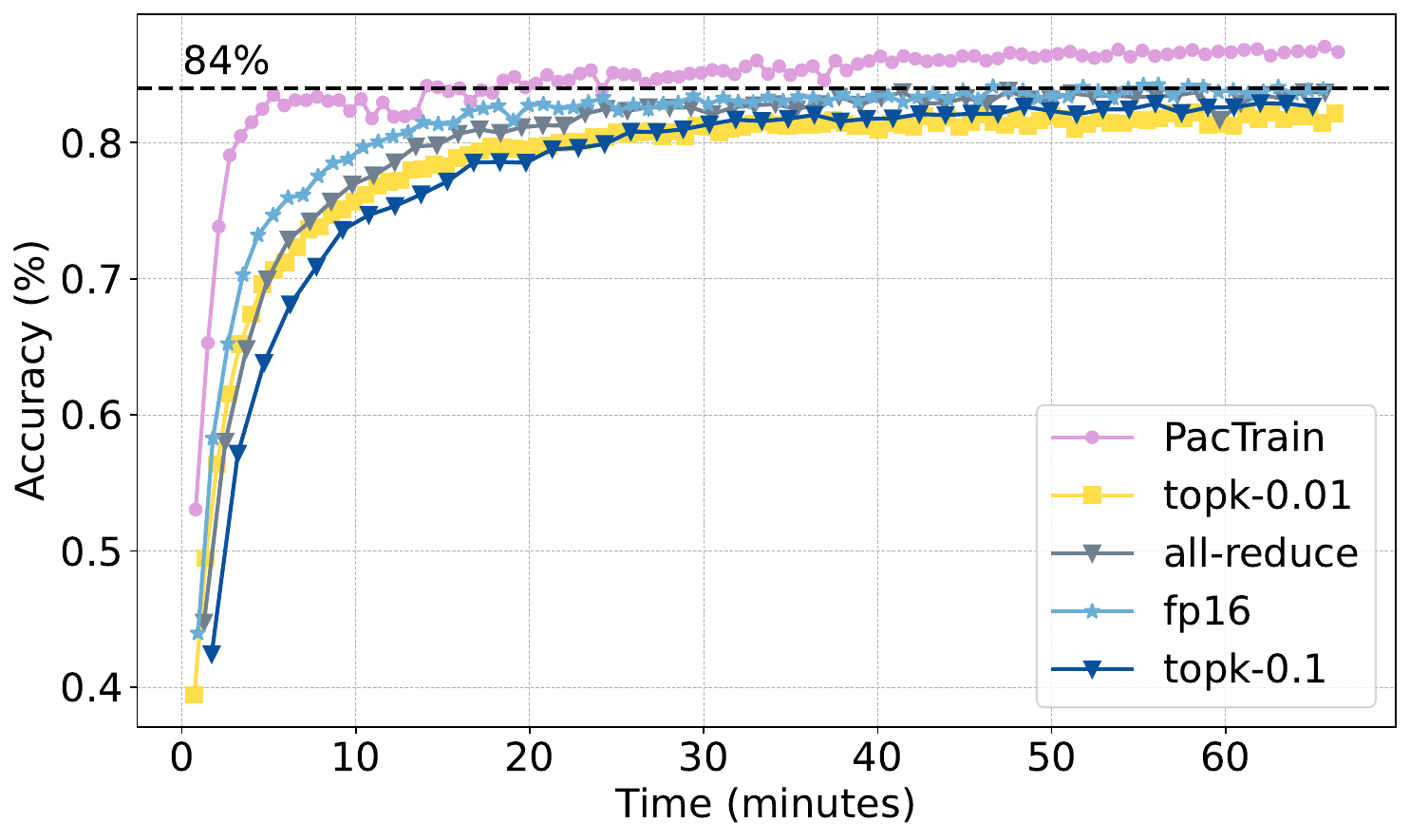}
    \caption{Time-to-accuracy Comparison of performance for the CIFAR-10 classification task using ResNet152.}
    \label{fig:resnet152}
\end{figure}

We compare ResNet152 on CIFAR-10 under a 1 Gbps bandwidth due to its representative slow convergence. Fig.~\ref{fig:resnet152} shows that for ResNet152, PacTrain reaches the 84\% target accuracy 5.64$\times$ faster than all-reduce baseline and 3.28$\times$ faster than fp16. For models where gradient components are evenly distributed, such as ResNet152, PacTrain tends to achieve a better convergence rate due to its utilization of pruning and quantization strategies. 

We observed that model compression algorithms like TopK, which are incompatible with all-reduce, can slow convergence under good network conditions. This is because TopK requires more epochs to reach the target accuracy. When gradient aggregation is not a bottleneck, the benefits of compression are minimal, potentially hindering convergence. TopK relies on all-gather for gradient aggregation, which can lead to accuracy loss with low compression ratios or excessive communication overhead with high compression ratios in distributed training. In contrast, PacTrain, being compatible with all-reduce, ensures communication cost scales are proportional to the pruning ratio.

\subsubsection{Impact of Pruning Ratio on Final Accuracy}

\begin{figure}
    \centering
    \includegraphics[width=0.8\linewidth]{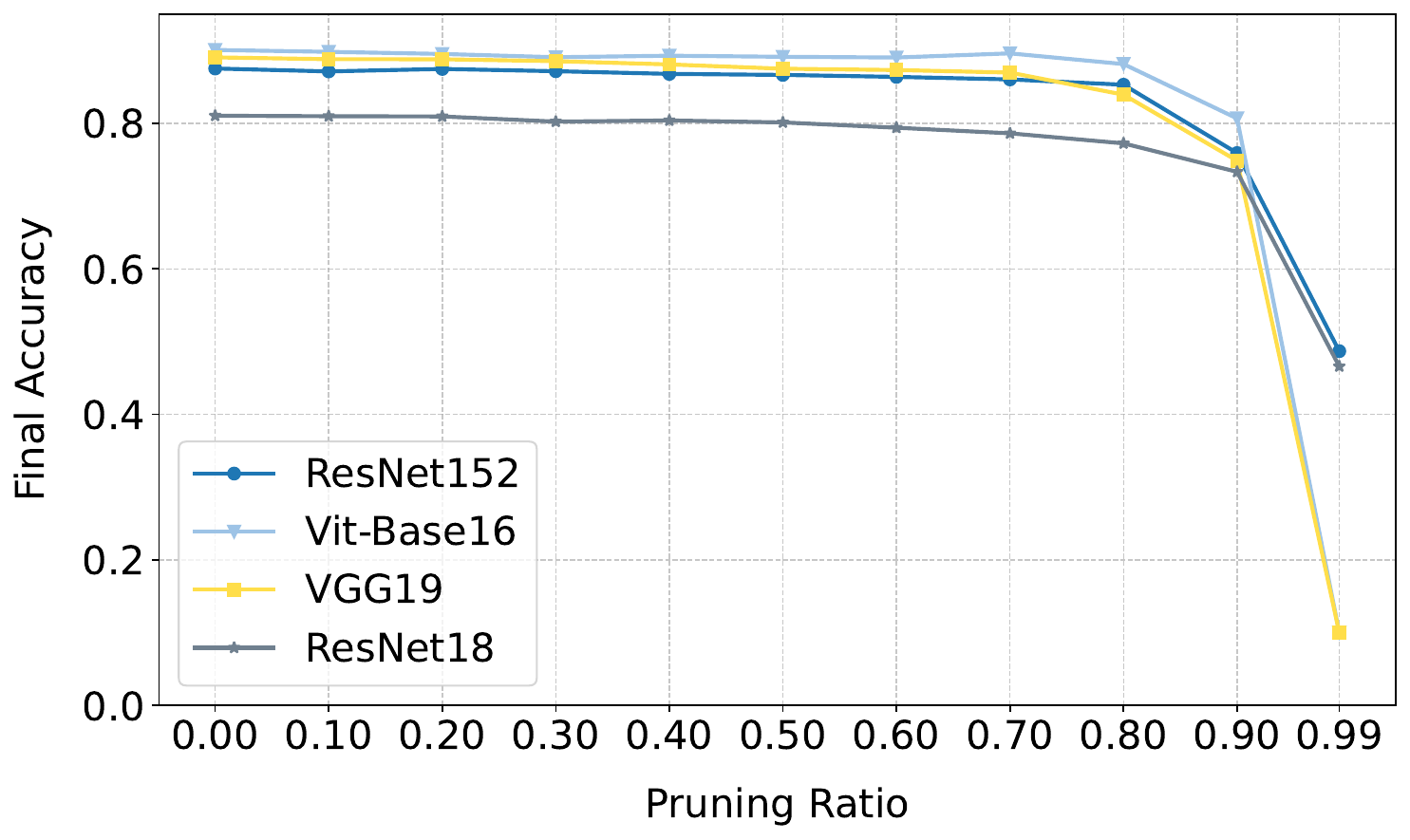}
    \caption{Impact of Pruning Ratio on Final Accuracy for CIFAR-10 classification task.}
    \label{fig:ratio_to_accuracy}
\end{figure}

Fig.~\ref{fig:ratio_to_accuracy} shows the impact of pruning ratios on final accuracy, based on our testing of VGG19, ResNet18, ResNet152, Vit-Base-16 models on the CIFAR-10 dataset. The relationship between the pruning ratio and final accuracy reflects a trade-off: as the pruning ratio increases, the final accuracy typically decreases due to the removal of important parameters. Although pruning introduces accuracy degradation, the impact remains minimal when the pruning ratio is kept below 80\%. Taking ResNet152 as an example, as long as the pruning ratio does not exceed 80\%, the final accuracy drops by less than 2\%.

\section{Conclusion}
\label{sec:conclusion}
This paper presents PacTrain, a framework that leverages sparse communication to accelerate pruning-aware distributed learning, offering a novel solution to the communication bottlenecks faced in the distributed training of large models. To demonstrate generalizability, we conducted several experiments to demonstrate that our method can effectively reduce communication costs and maintain the model accuracy across various training scenarios. Testbed experiments using eight GPU workers under various bandwidth-constrained network conditions show that PacTrain can achieve up to an 8.72× improvement.

\section{Acknowledgements}

This work is supported by the Guangdong Provincial Project 2023QN10X048, the Guangzhou Key Laboratory on Future Networked Systems 2024A03J0623, and the Guangzhou Science and Technology Project 2023A03J0011.

\newpage

\begingroup
\bibliographystyle{IEEEtran}
\footnotesize 
\bibliography{ref} 
\endgroup

\end{document}